\documentclass[aps,prl,reprint,showpacs,superscriptaddress,longbibliography,floatfix]{revtex4-1}
\usepackage{graphicx,mathrsfs,physics,amsmath,amssymb,bm}
\usepackage[colorlinks=true,linkcolor=blue,citecolor=blue,urlcolor=blue,hypertexnames=false]{hyperref}
\usepackage{url}

\begin{document}

\title{Gaussian-Bond Hall Viscosity and Nonlocal Hall Response of a Sampled Landau Level}

\author{Danyu Shu}
\email{shudanyu24@mails.ucas.ac.cn}
\affiliation{
Kavli Institute for Theoretical Sciences, University of Chinese Academy of Sciences, Beijing 100190, China
}

\author{Ryotaro Sano}
\affiliation{Institute of Solid State Physics, University of Tokyo, Kashiwa 277-8581, Japan}

\author{Ai Yamakage}
\affiliation{
Department of Physics, Nagoya University, Nagoya 464-8602, Japan
}

\author{Hiroshi Funaki}

\affiliation{
Center for Spintronics Research Network, Keio University, Yokohama 223-8522, Japan
}

\author{Mamoru Matsuo}
\email{mamoru@ucas.ac.cn}
\affiliation{
Kavli Institute for Theoretical Sciences, University of Chinese Academy of Sciences, Beijing 100190, China
}

\affiliation{
CAS Center for Excellence in Topological Quantum Computation, University of Chinese Academy of Sciences, Beijing 100190, China
}
\affiliation{
RIKEN Center for Emergent Matter Science (CEMS), Wako, Saitama 351-0198, Japan
}
\affiliation{
Advanced Science Research Center, Japan Atomic Energy Agency, Tokai, 319-1195, Japan
}

\date{\today}
\begin{abstract}
Hall viscosity on a lattice depends on how strain changes microscopic hopping amplitudes. For a specified Gaussian envelope, bond-length derivatives close on the undeformed hopping, expressing the stress as an embedded momentum Hessian and relating the viscosity to higher-order band geometry. In the Kapit--Mueller parent family at rational flux $\phi$, we show that the stress response equals the strain curvature of the sampled-LLL Gram matrix. The density-normalized viscosity approaches the continuum value $1/4$, with an exponentially small sampling correction at finite flux. By contrast, for a straight-bond Peierls source, a low-frequency mass, and a fixed-envelope susceptibility, the viscosity--conductivity comparison exhibits an algebraic deviation $-\phi/3+O(\phi^2)$, whose leading coefficient follows from a weak-field expansion.

\end{abstract}
\maketitle

\textit{Introduction.---}
In a continuum quantum Hall fluid, slow changes of spatial geometry produce a transverse stress proportional to the Hall viscosity, whose value in the isotropic quantum Hall states considered here is fixed by the orbital spin~\cite{j.e.avronViscosityQuantumHall1995,j.e.avronOddViscosity1998,n.readNonAbelianAdiabaticStatistics2009,carloshoyosHallViscosityElectromagnetic2012}. 
In a crystal the problem requires care, because the absence of continuous translation and rotation symmetry\cite{georgiosvarnavidesElectronHydrodynamicsAnisotropic2020} means that the response requires an explicit strain coupling: distinct microscopic deformations need not produce the same viscosity~\cite{hassanshapourianViscoelasticResponseTopological2015,thomasi.tuegelHallViscosityMomentum2015,taylorlhughesTorsionalResponseDissipationless2011,pranavraoHallViscosityQuantum2020}. Likewise, a finite-wave-vector electric response depends on the conserved current generated by the chosen Peierls source. Both ambiguities are particularly sharp for the Kapit--Mueller (K--M) model, whose occupied states are exact lattice samples of lowest-Landau-level (LLL) states~\cite{eliotkapitExactParentHamiltonian2010}.

We study deformations of microscopic hopping bonds, which underlie phonon Hall viscosity~\cite{maissambarkeshliDissipationlessPhononHall2012,b.flebusPhononHallViscosity2023}. Undeformed hopping amplitudes alone do not determine their bond-length derivatives. Specifying an analytic envelope supplies this missing deformation law for either short- or long-range hopping. For a Gaussian envelope, the derivatives close on the hopping amplitudes and give a quantum-geometric expression for Hall viscosity.

We evaluate the K--M viscosity both from magnetic-Bloch stress matrix elements and from the strain curvature of a sampled-LLL theta frame. The two expressions agree for the untruncated, metric-dependent parent Hamiltonian. The density-normalized viscosity approaches $1/4$, with an exponentially small correction arising from the discrete inner product in the Gram matrix. For specified electromagnetic and magnetic-field protocols, numerical calculations and a weak-field expansion instead give an algebraic deviation $-\phi/3+O(\phi^2)$ in the viscosity--conductivity comparison.

\textit{Bond stress.---}
The operator family $\hat H(u)$ specifies the mechanical source. We retain the symmetric part of the displacement gradient $u_{ij}=\partial_j u_i$, since the radial bond source does not couple linearly to a rigid rotation. The remaining strain components are a dilation and two shears. At a square reference point, $C_4$ symmetry forbids antisymmetric response between dilation and either shear, leaving the shear--shear coefficient considered here. We set the lattice spacing, hopping scale, and $\hbar$ to unity and parameterize the shears by

\begin{equation}
s_1=\frac12\begin{pmatrix}1&0\\0&-1\end{pmatrix},\quad
s_2=\frac12\begin{pmatrix}0&1\\1&0\end{pmatrix},\quad
u_{ij}=u_As_A^{ij}.
\label{eq:shear}
\end{equation}
When a Bloch decomposition is available, the Hall viscosity in our sign convention is the Brillouin-zone integral of the adiabatic shear curvature $\mathcal F_{12}=i\tr \hat{P}[\partial_{u_1}\hat{P},\partial_{u_2}\hat{P}]$, where $\hat{P}$ projects onto the occupied manifold~\cite{mvberryQuantalPhaseFactors,barrysimonHolonomyQuantumAdiabatic1983a}. A possible strain prescription is an affine change of momentum coordinates. It produces a second momentum moment of Berry curvature\cite{pranavraoHallViscosityQuantum2020,dingliElectronicHallViscosity2026}, but its lattice interpretation requires a prescription for the Brillouin-zone boundary and orbital embedding\footnote{We give a comparison between this affine Hall  viscosity and the bond Hall viscosity used here in continuum limit in the Supplemental Material}.

We instead specify the deformation of physical hopping bonds, as in phonon Hall viscosity~\cite{maissambarkeshliDissipationlessPhononHall2012,yunchaozhangPhononHallViscosity2021}, and keep the site labels fixed in the material frame. In a periodic representation, this prescription keeps the material-frame Bloch phase fixed and makes the strain coupling explicit. For orbital positions $\bm x_\alpha$, a hopping from orbital $\alpha$ to $\beta$ has physical displacement $\bm d=\bm R+\bm x_\beta-\bm x_\alpha$. Under $d_i\mapsto d_i+u_{ij}d_j$, the change of a radial envelope $t_{\alpha\beta}(d)$ is evaluated by a bond-dependent coefficient $\kappa_{\alpha\beta}(d)=-t'_{\alpha\beta}(d)/[d\,t_{\alpha\beta}(d)]$. Values of finitely many short-range hoppings do not fix these bond-length derivatives, so $\hat{H}_0$ alone generally does not determine the strain response. An explicitly specified analytic envelope fixes these derivatives across all bond lengths. In particular, a Gaussian hopping envelope $t(\bm d)\propto e^{-\kappa d^2/2}$ makes $\kappa_{\alpha\beta}(d)=\kappa$ independent of $\bm d$. The
first-order change of the hopping under a symmetric strain $u_{ij}$ is therefore $\delta t(\bm d)=-\kappa u_{ij}d_id_jt(\bm d)$, i.e. it is
governed by exactly the same $d_id_j$ factor that a momentum derivative of the hopping produces. This Gaussian derivative closure, rather than long range by itself, is the
decisive property. With this specified deformation law, the shear stress $\hat{T}_A=-\partial_{u_A}\hat{H}|_0$ becomes
\begin{equation}
\hat{T}_A=-\kappa s_A^{ij}\mathcal D_i\mathcal D_j \hat{H},
\qquad \mathcal D_i\hat O=-i[\hat r_i,\hat O].
\label{eq:stress}
\end{equation}
Here $\mathcal D_i$ acts on a real-space hopping $|\bm R\rangle\langle\bm R+\bm d|$ as multiplication by $id_i$, including the physical orbital displacement. Where a Bloch representation exists, this commutator becomes the embedded momentum derivative~\cite{ohOrbitalEmbeddingPhysical2026,fuchsOrbitalEmbeddingTopology2021}. Its real-space definition does not require rational magnetic flux. An area-preserving deformation preserves the magnetic cocycle. With the material-frame gauge connection transformed consistently, the stress follows from the envelope derivative, and Eq.~\eqref{eq:stress} remains valid at nonzero magnetic field.

\textit{Hall viscosity and quantum geometry.---}
For an isolated occupied manifold admitting a Bloch decomposition, the signed shear response in our orientation is
\begin{equation}
\eta_H=-2\int_{\text{BZ}}\!\frac{\mathrm{d}^2k}{(2\pi)^2}
\operatorname{Im}\!\sum_{n\in\mathrm{occ},m\in\mathrm{emp}}
\frac{(\hat{T}_1)_{nm}(\hat{T}_2)_{mn}}{(E_m-E_n)^2}.
\label{eq:kubo}
\end{equation}
Here $\hat{P}$ and $\hat{Q}=1-\hat{P}$ project onto the occupied and
empty manifolds. Because $\hat{T}_A$ has the dimensions of energy, $\eta_H$
has the dimensions of density; we therefore report the dimensionless density-normalized quantity $\bar\eta_H=\eta_H/(C\rho)$,
where $C$ is the Chern number and $\rho$ the particle density.
This normalization gives $\bar\eta_H = 1/4$ for a filled continuum LLL, which is the reference value used below.

Equation~\eqref{eq:stress} yields an explicit geometric expression. Consider a two-band model with $\hat{H}=E \hat{P}+(E+\Delta)\hat{Q}$, where $E$ and $\Delta$ are the occupied energy and positive gap. Let $\hat{\rho}_i=\hat{Q}(\mathcal D_i\hat{P})\hat{P}$, $\hat{\Gamma}_{ij}=\hat{Q}(\mathcal D_i\mathcal D_j\hat{P})\hat{P}$, $w_i=\partial_{k_i}\ln\Delta$, and $\hat{M}_{ij}=\hat{\Gamma}_{ij}+w_i\hat{\rho}_j+w_j\hat{\rho}_i$~\cite{mitscherlingGaugeinvariantProjectorCalculus2025,ahnRiemannianGeometryResonant2022}.
The Hall viscosity in Eq.~\eqref{eq:kubo} depends on the stress matrix elements between occupied and empty states:
\begin{equation}
\hat{Q}\hat{T}_{ij}\hat{P}=\kappa\Delta \hat{M}_{ij}.
\label{eq:two-band}
\end{equation}
The explicit gap factors cancel in Eq.~\eqref{eq:kubo}, leaving a bilinear in covariant projector derivatives and logarithmic gap gradients. In particular, $\tr(\hat{\rho}_i^\dagger\hat{\rho}_j)=g_{ij}-i\Omega_{ij}/2$ defines the quantum metric $g_{ij}$ and signed Berry curvature $\Omega_{ij}$ in this convention~\cite{j.p.provostRiemannianStructureManifolds1980,tomokiozawaRelationsTopologyQuantum2021}. Define $\mathcal T_{ijk}=\operatorname{Im}(c_{ij;k}+c_{jk;i}+c_{ki;j})/3$, with $c_{ij;k}=\tr(\hat{\Gamma}_{ij}^\dagger\hat{\rho}_k)$. Before the shear contraction, the $xx|xy$ integrand $\mathcal I_{xx|xy}:=\operatorname{Im}\tr(\hat{M}_{xx}^\dagger \hat{M}_{xy})$ has the exact separation
\begin{align}
\mathcal I_{xx|xy}&=\mathcal K^{(4)}_{xx|xy}
-2\mathcal B_{xx|xy}-w_x^2\Omega_{xy},\nonumber\\
\mathcal B_{xx|xy}&=\frac{w_x}{3}\partial_x\Omega_{xy}
+\frac{w_x\mathcal T_{xxy}-w_y\mathcal T_{xxx}}{2},
\label{eq:two-band-geometry}
\end{align}
where $\mathcal K^{(4)}_{xx|xy}=\operatorname{Im}\tr(\hat{\Gamma}_{xx}^\dagger\hat{\Gamma}_{xy})$. The three terms are respectively a fourth-order invariant, a third-order connection term, and a quantum-geometric-tensor (QGT) term weighted by the gap gradient.\footnote{The Supplemental Material derives this decomposition and reconstructs the fourth-order invariant from rank-three data wherever the first projector derivatives do not all vanish.} In particular, the same $\mathcal B_{xx|xy}$ term appears in the nonlocal Hall conductivity of Kozii \emph{et al.}~\cite{vladyslavkoziiIntrinsicAnomalousHall2021}, where the full response also contains other geometric and band-energy terms.

This result connects the Berry curvature in strain space to momentum-space projector geometry for a specified Gaussian deformation. The shared geometric structures provide a basis for comparing viscoelastic and electromagnetic responses while keeping their distinct source couplings and spectral weights explicit.

\textit{Exact K--M viscosity.---}
A particular Gaussian model is the Kapit--Mueller model, whose ground-state manifold realizes a lattice-sampled LLL with Chern number $C=-1$. It provides a natural benchmark for determining how lattice sampling modifies the continuum Hall viscosity. 

On the infinite square lattice, let $B=2\pi\phi$ with $0<\phi<1$. In Landau gauge $\bm A_0=(0,Bx)$, define
\begin{equation}
\hat{\mathcal W}_{\bm d}
=\sum_{\bm R}e^{-iBd_y(R_x+d_x/2)}
|\bm R\rangle\langle\bm R+\bm d|.
\label{eq:km-translation}
\end{equation}
The K--M Hamiltonian is $\hat H=\sum_{\bm d\ne0}t(\bm d)\hat{\mathcal W}_{\bm d}$ where the hopping strength $t(\bm d)$ is give by
\begin{align}
t(\bm d)=\varsigma_{\bm d}e^{-\pi(1-\phi)|\bm d|^2/2},
\quad \varsigma_{\bm d}=(-1)^{d_x+d_y+d_xd_y}.
\label{eq:km-hop}
\end{align}
These real-space operators are defined for arbitrary flux, including irrational $\phi$. They obey $\mathcal D_i\hat{\mathcal W}_{\bm d}=id_i\hat{\mathcal W}_{\bm d}$, so the same commutator derivative generates the bond stress and the uniform current.

Equivalently, strain can be parameterized by a positive unimodular inverse metric $\gamma^{ij}$ that deforms the Gaussian through $|\bm d|^2\mapsto \gamma^{ij}d_id_j$. At rational flux $\phi=p/q$, the generalized coherent-state sum rule gives an exact rank-$p$ parent eigenspace in each magnetic Bloch fiber for this entire metric family, provided the hopping is untruncated and the occupied manifold remains isolated~\cite{perelomovGeneralizedCoherentStates1977,shenExactParentHamiltonians2026}. 

Parameterize the metric by
\begin{equation}
\gamma_{ij}(\tau)=\frac1{\tau_2}
\begin{pmatrix}1&\tau_1\\\tau_1&|\tau|^2\end{pmatrix},
\qquad \tau=\tau_1+i\tau_2.
\label{eq:metric}
\end{equation}
We choose the convention that its variation near $\tau=i$ reproduces the bond strain through $u_1=\delta\tau_2$ and $u_2=-\delta\tau_1$. The generalized Poisson summation identity then gives the exact ground-state manifold of the strained K--M parent Hamiltonian. In Landau gauge, a basis for this manifold is provided by the sampled-LLL magnetic-Bloch frame, with $q\times p$ matrix elements~\cite{haldaneManyParticleTranslationalSymmetries1985,haldanePeriodicLaughlinJastrowWave1985} 

\begin{widetext}
\begin{equation}
\Phi_{ar}(\bm k;\tau)=e^{-ia k_x}\sum_{n\in\mathbb Z}e^{inqk_x}
\exp\!\left[\frac{i\pi\tau p}{q}
\left(a-\frac{qk_y}{2\pi p}-\frac{qr}{p}-nq\right)^2\right],
\quad a=0,\ldots,q-1,\quad r=0,\ldots,p-1.
\label{eq:theta}
\end{equation}
\end{widetext}

Define the Gram matrix~\cite{youjiangxuBuildingFlatbandLattice2020,carolinapaivaGeometricalResponsesGeneralized2025,liuTheoryGeneralizedLandau2025}
$G=\Phi^\dagger\Phi$, the occupied projector
$\hat{P}=\Phi G^{-1}\Phi^\dagger$, and
$\hat{Q}=1-\hat{P}$.
The holomorphicity condition $\partial_{\bar\tau}\Phi=0$
endows the occupied manifold with ideal quantum geometry in
strain space. At the square reference point $\tau=i$, it implies
the chiral stress selection rule
$\hat{Q}(\hat{T}_1+i\hat{T}_2)\hat{P}=0$ and allows the
viscosity to be expressed directly through the Gram matrix
\footnote{See Supplemental Material for a detailed derivation}. With
$M_\tau=(\partial_\tau\Phi)^\dagger\hat{Q}\partial_\tau\Phi$,
\begin{equation}
\bar\eta_H
=\frac{2}{p}\left\langle\tr(G^{-1}M_\tau)\right\rangle_{\text{MBZ}}
=\frac{2}{p}\left\langle
\partial_{\bar\tau}\partial_\tau\ln\det G
\right\rangle_{\text{MBZ}},
\label{eq:gram}
\end{equation}
where all quantities are evaluated at $\tau=i$.
Here $\rho=B/(2\pi)=\phi$ is the particle density, and
$\langle f\rangle_{\text{MBZ}}
=q\int_{\text{MBZ}}\mathrm{d}^2k\,f/(2\pi)^2$
denotes the uniform magnetic-BZ average.

\begin{figure}[t]
\includegraphics[width=\columnwidth]{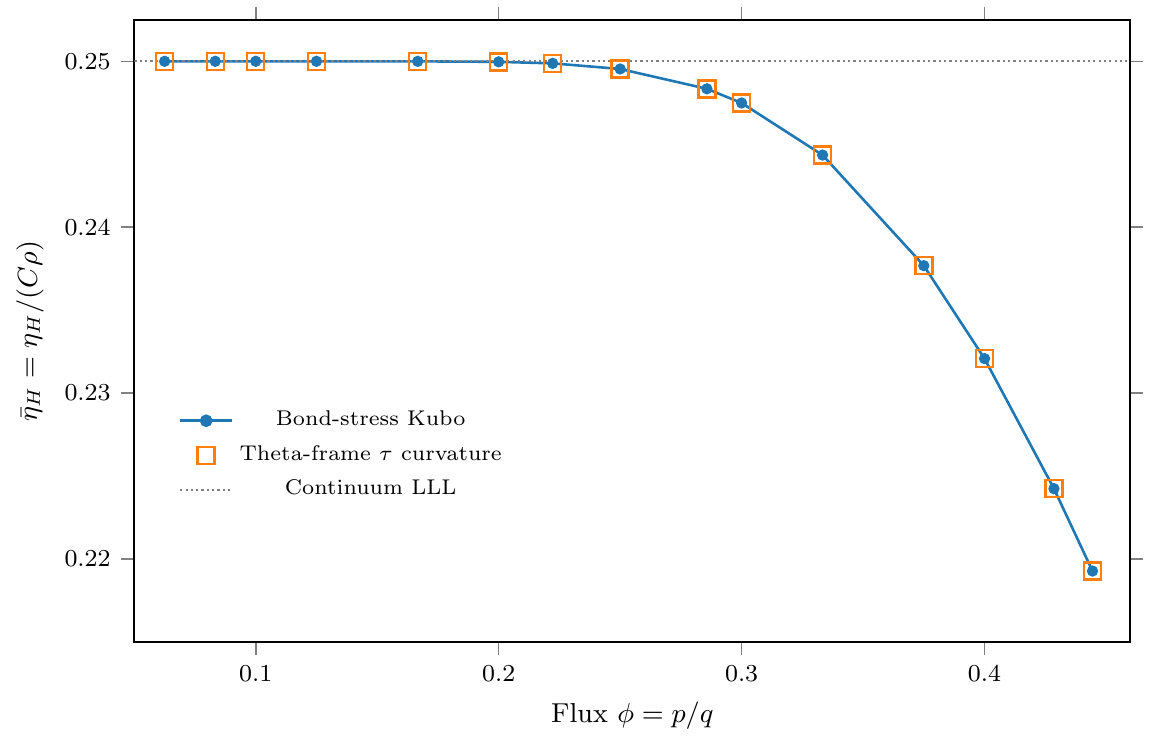}
\caption{Density-normalized K--M bond Hall viscosity
$\bar\eta_H$. Circles show the Gaussian bond-stress Kubo results,
open squares show the theta-Gram results, and the dotted line
marks the continuum LLL value.}
\label{fig:viscosity}
\end{figure}

The Gram matrix encodes the normalization and mutual overlaps
of the sampled states. For a holomorphic frame, the mixed
derivative of $\ln\det G$ gives the curvature of the occupied
subspace and is invariant under nonsingular holomorphic changes
of occupied basis. Lattice sampling changes the strain dependence
of these overlaps and hence the viscosity, while preserving the
integer Chern number. The weak-flux Poisson-image expansion of
Eq.~\eqref{eq:gram}, with $x=\pi/\phi$, yields
\begin{equation}
\bar\eta_H=\frac14-(x^2-2x)e^{-x}
+O(x^2e^{-3x/2}).
\label{eq:asymptotic}
\end{equation}
The zero Poisson image reproduces the continuum value $1/4$,
whereas nonzero images give the lattice-sampling correction.
Physically, the lattice Gram matrix sums overlaps over discrete
sites, while its continuum counterpart integrates them over
space; their different strain dependence produces
Eq.~\eqref{eq:asymptotic}.
Figure~\ref{fig:viscosity} compares the stress Kubo calculation
with the exact theta-Gram expression in Eq.~\eqref{eq:gram}.

The sampled-LLL structure also gives the occupied manifold
ideal quantum geometry in momentum space~\cite{brunomeraKahlerGeometryChern2021,patrickgallagherQuantumcriticalConductivityDirac2019}.
The identity
$\hat{Q}\partial_{k_y}\Phi
=\tau\hat{Q}\partial_{k_x}\Phi$
fixes the quantum metric of the full rank-$p$ occupied manifold:
\begin{equation}
g_{ij}(\bm k;\tau)
=\frac{|\Omega_{xy}(\bm k;\tau)|}{2}\gamma_{ij}(\tau).
\label{eq:ideal-metric}
\end{equation}
This relation implies the generalized trace condition
$\gamma^{ij}g_{ij}=|\Omega_{xy}|$ throughout the isolated
parent family. At $\tau=i$, it reduces to the Euclidean
trace condition; for general $\tau$, the ideal geometry is
adapted to the anisotropic metric $\gamma_{ij}$ and does not
require uniform Berry curvature. This momentum-space property
characterizes the model's LLL geometry, whereas
Eq.~\eqref{eq:gram} follows specifically from holomorphicity
in the strain parameter.

\textit{Nonlocal Hall response.---}
In a Galilean-invariant continuum, a stress--current Ward identity relates Hall viscosity, the quadratic Hall-conductivity coefficient, and magnetic susceptibility~\cite{carloshoyosHallViscosityElectromagnetic2012,barrybradlynKuboFormulasViscosity2012,biaohuangHallViscosityRevealed2015,lucav.delacretazTransportSignaturesHall2017}. We compare this expression with the K--M response for explicit current and magnetic-field protocols.

Write the single-particle site-density operator as
$\hat N(\bm q)=e^{-i\bm q\cdot\hat{\bm r}}$
Integrating the probe vector potential along each straight bond gives
$\hat J_i(\bm q)=-\delta\hat H[A]/\delta A_i(-\bm q)|_{A=0}$,
with single-particle kernel
\begin{align}
\hat J_i(\bm q)
&=i\sum_{\bm d\ne0}d_i t(\bm d)
\operatorname{sinc}(\bm q\cdot\bm d/2)\nonumber\\
&\quad\times\hat N(\bm q/2)\hat{\mathcal W}_{\bm d}\hat N(\bm q/2),
\label{eq:current-line}
\end{align}
where $\operatorname{sinc}(z)=\sin z/z$. Its fermion bilinear is the physical current. The exact charge Ward identity is
\begin{equation}
q_i\hat J_i(\bm q)=[\hat N(\bm q),\hat H],
\label{eq:charge-ward}
\end{equation}
which gives $\dot{\hat N}(\bm q)+iq_i\hat J_i(\bm q)=0$ and the corresponding many-body continuity equation. At $\bm q=0$, $\hat J_i(0)=\mathcal D_i\hat H$. Neither this definition nor the identity requires rational flux or a Bloch Hamiltonian. Charge conservation constrains the longitudinal current, while the Peierls path fixes its transverse component. Paths with the same endpoints can therefore produce different finite-wave-vector responses while satisfying the same continuity equation.

For $\bm q=q_x\bm e_x$, write the static Hall expansion as
\begin{equation}
\frac{\sigma_H(q_x)}{\sigma_H(0)}
=1+C_2\frac{q_x^2}{B}+O(q_x^4).
\label{eq:c2}
\end{equation}

\begin{figure}[tbp]
\includegraphics[width=\columnwidth]{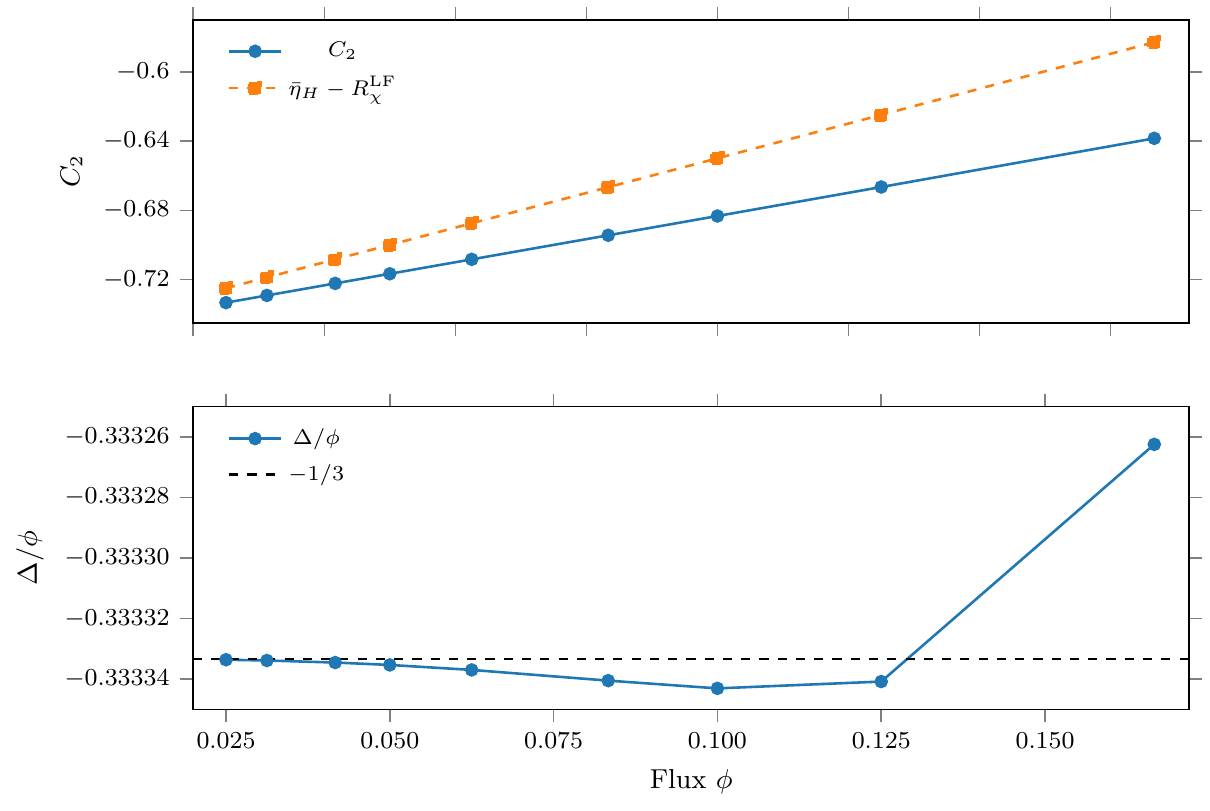}
\caption{Hall response and lattice correction. (Top) The Hall coefficient $C_2$ of the conserved straight-bond Peierls current and the independently computed viscosity--susceptibility expression. (Bottom) Their deviation $\Delta/\phi$; the dashed line marks the derived weak-flux coefficient $-1/3$. }
\label{fig:hoyos}
\end{figure}
The continuum comparison $\bar\eta_H-R_\chi^{\rm LF}$~\cite{carloshoyosHallViscosityElectromagnetic2012,mohammadsherafatiHallViscosityElectromagnetic2016,harperFinitewavevectorElectromagneticResponse2018} uses $R_\chi^{\rm LF}=-2\pi m_{\rm LF}\chi_B$, with magnetic susceptibility and low-frequency mass defined by
\begin{align}
\chi_B
&=-\left.\partial_B^2\mathcal E\right|_{\kappa,\gamma,\nu=1},\nonumber\\
\sigma_{xx}(\omega)&=-i\omega\frac{\rho m_{\rm LF}}{B^2}
+O(\omega^3),\qquad \rho=\frac{B}{2\pi},
\label{eq:susceptibility-definition}
\end{align}
Here $\mathcal E$ is the energy per reference area. The field derivative holds the Gaussian envelope $\kappa$ and geometry $\gamma$ fixed at filling $\nu=1$; density varies as $B/(2\pi)$. It is evaluated at each parent point, along a path distinct from $\kappa=\pi(1-\phi)$. The low-frequency mass is a specified response coefficient; identifying it with a unique lattice inertial mass is not assumed.

Write the correction to the continuum expression as
\begin{equation}
\Delta(\phi):=C_2(\phi)-[\bar\eta_H(\phi)-R_\chi^{\rm LF}(\phi)].
\label{eq:defect}
\end{equation}
The leading lattice correction can be obtained analytically
from a weak-field expansion~\cite{harperFinitewavevectorElectromagneticResponse2018,fennerharperPerturbativeApproachFlat2014}.
Representing the magnetic hopping operators as
$\hat{\mathcal{W}}_{\bm d}\mapsto e^{i\bm d\cdot\hat{\bm\pi}}$, with $[\hat{\pi}_x,\hat{\pi}_y]=iB$,
we expand the K--M Hamiltonian along its parent line
$\kappa=\pi-B/2$. Through fourth order in covariant momentum,
\begin{equation}
\hat{H}_{\rm eff}
=-1+\alpha(\hat{\bm\pi}^2-B)
+\beta(\hat{\bm\pi}^2-B)(\hat{\bm\pi}^2-3B)+O(B^3).
\label{eq:weak-effective}
\end{equation}
Here $\alpha$ and $\beta$ are determined by the zero-field
Gaussian hopping moments, whose Perelomov identities fix
$\beta/\alpha=-1/(4\pi)$.
The remainder is counted at fixed Landau-level index,
where $\hat{\bm\pi}^2=O(B)$ since $\hat{\bm\pi}^2\ket{n}=B(2n+1)\ket{n}$; the ordering of the noncommuting
momenta is inherited from the magnetic-translation
exponentials.

Keeping the same straight-bond Peierls source, the
density--current response through cubic order in wave
vector involves the first two oscillator excitations,
whereas the uniform low-frequency mass probes only the
first. Their contributions give
$C_2=-3/4+2\phi/3+O(\phi^2)$ and
$m_{\rm LF}=1/(2\alpha)+O(B^2)$.
Varying the magnetic field at fixed Gaussian envelope,
rather than along the parent line, then gives
$R_\chi^{\rm LF}=1-\phi+O(\phi^2)$.
Together with the Gram-matrix result
$\bar\eta_H=1/4$ to algebraic order, these yield
\begin{equation}
\Delta
=C_2-\bar\eta_H+R_\chi^{\rm LF}
=-\frac{\phi}{3}+O(\phi^2).
\label{eq:weak-hoyos}
\end{equation}
The response limits are taken at fixed $B>0$ before
$\phi\to0$, and only the linear correction is controlled
by this truncation. The result agrees with the lower
panel of Fig.~\ref{fig:hoyos}.

\textit{Conclusion.---}
For a specified Gaussian bond deformation, the stress is an embedded momentum Hessian, exposing higher-order projector geometry and logarithmic gap gradients in the two-band Hall response. In the K--M parent family, its viscosity equals the theta-Gram strain curvature, with an exponentially small sampling correction to $\bar\eta_H=1/4$. The specified electromagnetic response instead departs algebraically from the continuum viscosity--conductivity expression. This contrast separates the strain geometry of the occupied states from the additional spectral and source information entering the nonlocal Hall response.

\begin{acknowledgments}
R.S. is grateful to Yuma Hirobe and Taisei Kitamura for fruitful discussions. This work was supported by the National Natural Science Foundation of China (NSFC) under Grant No. 12374126, by the Priority Program of the Chinese Academy of Sciences under Grant No. XDB28000000, by the Hirose Foundation, and by JSPS KAKENHI Grant Nos. JP23H01839, JP24H00322, JP24K06951, JP24H00853, and JP25K07224.
\end{acknowledgments}

\bibliography{ref}

\clearpage\onecolumngrid
\begin{center}\textbf{Supplemental Material for ``Gaussian-Bond Hall Viscosity and Nonlocal Hall Response of a Sampled Landau Level''}\end{center}
\setcounter{equation}{0}\renewcommand{\theequation}{S\arabic{equation}}\renewcommand{\theHequation}{S\arabic{equation}}
\setcounter{figure}{0}\renewcommand{\thefigure}{S\arabic{figure}}\renewcommand{\theHfigure}{S\arabic{figure}}

\section{Affine and bond deformation in the continuum limit}

An affine change of reciprocal coordinates defines
$\hat{H}_{\rm aff}(\bm k,u)=\hat{H}[(1-u^T)\bm k]$ and the corresponding occupied projector. Thus $\partial_{u_A}\hat{P}=-s_A^{ij}k_j\partial_{k_i}\hat{P}$, giving
\begin{equation}
\eta^{\rm aff}_{AB}=\int_{\rm BZ}\frac{d^2k}{(2\pi)^2}
(s_A\bm k)^i(s_B\bm k)^j\Omega_{ij}(\bm k),
\label{eq:affine-moment}
\end{equation}
where $\Omega_{ij}=i\tr \hat{P}[\partial_{k_i}\hat{P},\partial_{k_j}\hat{P}]$ is the Berry curvature~\cite{mvberryQuantalPhaseFactors,barrysimonHolonomyQuantumAdiabatic1983a}.
The explicit momentum factors require a prescription for the Brillouin-zone boundary and orbital embedding: the bare integrand is not periodic on a fixed momentum torus.

The continuum relation to bond deformation is transparent at zero magnetic field. For a scalar hopping envelope,
$\hat{H}(\bm k)=\sum_{\bm d}t(\bm d)e^{i\bm k\cdot\bm d}$, the two stresses are
\begin{equation}
\hat{T}_{ij}^{\rm aff}=k_i\partial_{k_j}\hat{H},\qquad
\hat{T}_{ij}^{\rm bond}=-\sum_{\bm d}d_j\partial_{d_i}t(\bm d)e^{i\bm k\cdot\bm d}.
\end{equation}
Replacing the sum by a continuum integral, sufficiently rapid decay of $t$ removes the boundary term and gives
\begin{equation}
-\int d^2d\,d_j\partial_{d_i}t(\bm d)e^{i\bm k\cdot\bm d}
=\delta_{ij}\int d^2d\,t(\bm d)e^{i\bm k\cdot\bm d}
+ik_i\int d^2d\,d_jt(\bm d)e^{i\bm k\cdot\bm d}.
\end{equation}
Hence $\hat{T}_{ij}^{\rm bond}=\hat{T}_{ij}^{\rm aff}+\delta_{ij}\hat{H}$ in this continuum representation. The extra term drops out for traceless strain, so the shear stresses coincide. Moreover, for the adiabatic Berry-curvature response of a gapped noninteracting system, the additional term has no occupied–unoccupied matrix elements and therefore does not modify the strain-space Berry curvature.

On a Bravais lattice of primitive-cell area, Poisson summation instead yields
\begin{align}
\hat{T}_{ij}^{\rm bond}
&=\hat{T}_{ij}^{\rm aff}+\delta_{ij}\hat{H}+\Delta_{ij}^{\rm lat},\nonumber\\
\Delta_{ij}^{\rm lat}(\bm k)
&=\frac1{V_c}\sum_{\bm G}G_i\partial_{k_j}\widetilde t(\bm k+\bm G),\qquad
\widetilde t(\bm k)=\int d^2r\,t(\bm r)e^{i\bm k\cdot\bm r}.
\end{align}
Here $\bm G$ runs over reciprocal-lattice vectors. These expressions assume the indicated smooth interpolation of the hopping and convergence of the sums and derivatives. At nonzero magnetic field, a physical affine transformation must also transform the magnetic connection; a bare substitution of Bloch momentum alone does not establish equivalence to bond deformation.

\section{Two-band quantum-geometric contraction}

For $\hat{H}=E\mathbb I+\Delta \hat{Q}$, where $\hat{Q}=1-\hat{P}$ and $\Delta>0$, define
$\hat{\rho}_i=\hat{Q}(\mathcal D_i\hat{P})\hat{P}$, $\hat{\Gamma}_{ij}=\hat{Q}(\mathcal D_i\mathcal D_j\hat{P})\hat{P}$,
and $w_i=\partial_i\ln\Delta$.
The Gaussian stress $\hat{T}_{ij}=-\kappa\mathcal D_i\mathcal D_j\hat{H}$ obeys
\begin{equation}
 \hat{Q}\hat{T}_{ij}\hat{P}=\kappa\Delta \hat{M}_{ij},\qquad
 \hat{M}_{ij}=\hat{\Gamma}_{ij}+w_i\hat{\rho}_j+w_j\hat{\rho}_i,
 \qquad
 \eta^H_{AB}=-2\kappa^2\int_{\rm BZ}\frac{d^2k}{(2\pi)^2}
 \operatorname{Im}\tr(\hat{M}_A^\dagger \hat{M}_B),
 \label{eq:s-two-band-kubo}
\end{equation}
where $\hat{M}_A=s_A^{ij}\hat{M}_{ij}$. This follows by differentiating
$\hat{H}=E\mathbb I+\Delta \hat{Q}$: terms proportional to $\mathbb I$ or $\hat{Q}$
vanish in its $\hat{Q}\hat{P}$ block, and the Kubo denominator cancels $\Delta^2$.

For rank-one $\hat{P}$ and $\hat{Q}$, write
$\hat{\rho}_i=|v\rangle a_i\langle u|$ and
$\hat{\Gamma}_{ij}=|v\rangle b_{ij}\langle u|$.
In an embedded local frame,
$a_i=\langle v|\partial_i u\rangle$ and
$b_{ij}=(\partial_i+A_i^v-A_i^u)a_j$,
with $A_i^u=\langle u|\partial_i u\rangle$ and $A_i^v=\langle v|\partial_i v\rangle$.
Commuting embedded derivatives give $b_{ij}=b_{ji}$. Set
\begin{equation}
 a_i^*a_j=g_{ij}-\frac{i}{2}\Omega_{ij},\qquad
 c_{ij;k}=b_{ij}^*a_k,\qquad
 \mathcal T_{ijk}=\frac13\operatorname{Im}
 (c_{ij;k}+c_{jk;i}+c_{ki;j}).
 \label{eq:s-geometric-data}
\end{equation}
Differentiating $a_i^*a_j$ yields
\begin{align}
 2\operatorname{Re}c_{ij;k}
 &=\partial_i g_{jk}+\partial_j g_{ik}
   -\partial_k g_{ij},\nonumber\\
 \partial_x\Omega_{xy}
 &=-2\operatorname{Im}(c_{xx;y}-c_{xy;x}).
 \label{eq:s-geometric-derivatives}
\end{align}
Writing $\hat{M}_{ij}=|v\rangle m_{ij}\langle u|$, with $m_{ij}=b_{ij}+w_i a_j+w_j a_i$, and using these identities gives
\begin{equation}
 \operatorname{Im}(m_{xx}^*m_{xy})
 =\mathcal K^{(4)}_{xx|xy}
 -\frac23w_x\partial_x\Omega_{xy}
 -w_x\mathcal T_{xxy}+w_y\mathcal T_{xxx}
 -w_x^2\Omega_{xy},
 \qquad
 \mathcal K^{(4)}_{xx|xy}=\operatorname{Im}(b_{xx}^*b_{xy}).
 \label{eq:s-two-band-geometry}
\end{equation}
The shear integrand is one half of the difference between the
$xx|xy$ and $yy|xy$ contractions. Where $\tr g>0$,
\begin{equation}
 \mathcal K^{(4)}_{xx|xy}
 =\frac{\operatorname{Im}(c_{xx;x}c_{xy;x}^*
                         +c_{xx;y}c_{xy;y}^*)}{\tr g}.
 \label{eq:s-fourth-from-third}
\end{equation}
At zeros of the first derivatives of $\hat{P}$, its nonsingular definition in
Eq.~\eqref{eq:s-two-band-geometry} must be used.
The $\partial_x\Omega$--$\mathcal T$ combination also appears in the
two-band conductivity of Kozii \emph{et al.}~\cite{vladyslavkoziiIntrinsicAnomalousHall2021}.
The shared geometric terms do not equate the full responses, which
contain different contractions and require specified source vertices.

\section{Sampled LLL states and the Gram-matrix curvature}

Parameterize the unimodular metric by
\begin{equation}
 \gamma_{ij}(\tau)=\frac1{\tau_2}
 \begin{pmatrix}1&\tau_1\\\tau_1&|\tau|^2\end{pmatrix},
 \qquad z_\tau(\bm r)=\frac{\tau x-y}{\sqrt{\tau_2}},
 \qquad |z_\tau(\bm d)|^2=d_i\gamma^{ij}d_j.
 \label{eq:s-metric}
\end{equation}
The generalized Perelomov identity is
\begin{equation}
 \sum_{\bm d\in\mathbb Z^2}\varsigma_{\bm d}
 e^{-\pi|z_\tau(\bm d)|^2/2}F(z_\tau(\bm d))=0,
 \qquad \varsigma_{\bm d}=(-1)^{d_x+d_y+d_xd_y},
 \label{eq:s-perelomov}
\end{equation}
for entire $F$ for which the sum converges. It follows from Poisson
summation on the unit-area lattice transformed by $z_\tau$.
At $\kappa=\pi-B/2$, include the on-site term by writing $\hat{K}=\hat{H}+\mathbb I$.
For a sampled LLL function
$\psi_f(\bm r)=e^{-B|z_\tau(\bm r)|^2/4}f(z_\tau(\bm r))$,
the symmetric-gauge kernel carries the phase $e^{-iB\bm r\times\bm d/2}$ and gives
\begin{equation}
 (\hat{K}\psi_f)(\bm r)
 =e^{-B|z_\tau(\bm r)|^2/4}
 \sum_{\bm d}\varsigma_{\bm d}e^{-\pi|z_\tau(\bm d)|^2/2}
 e^{-B\overline{z_\tau(\bm r)}z_\tau(\bm d)/2}
 f(z_\tau(\bm r)+z_\tau(\bm d))=0.
 \label{eq:s-parent-zero-mode}
\end{equation}
The exponent uses
$\operatorname{Im}[\overline{z_\tau(\bm r)}z_\tau(\bm d)]
=\bm r\times\bm d$; the remaining summand is precisely of the form
in Eq.~\eqref{eq:s-perelomov}. Thus these states retain energy $-1$
under the metric deformation, on the gapped parent branch.

In Landau gauge, the corresponding orbitals are
$e^{ik_y^\prime y}\exp[iB\tau(x-k_y^\prime/B)^2/2]$.
For $\phi=p/q$, magnetic Bloch sums with
$k_y^\prime=k_y+2\pi(r+pn)$ give the $q\times p$ frame, with
$a=0,\ldots,q-1$ and $r=0,\ldots,p-1$,
\begin{equation}
 \Phi_{ar}(\bm k;\tau)
 =e^{-iak_x}\sum_{n\in\mathbb Z}e^{inqk_x}
 \exp\!\left[i\pi\tau\frac pq
 \left(a-\frac{qk_y}{2\pi p}-\frac{qr}{p}-nq\right)^2\right].
 \label{eq:s-theta-frame}
\end{equation}
Consequently $\hat{K}\Phi=0$. Let $G=\Phi^\dagger\Phi$ and
$\hat{P}=\Phi G^{-1}\Phi^\dagger$.
The frame is holomorphic in $\tau$, so differentiation gives
\begin{equation}
 \partial_{\bar\tau}\partial_\tau\ln\det G
 =\tr\!\left[G^{-1}(\partial_\tau\Phi)^\dagger
 \hat{Q}\partial_\tau\Phi\right],\qquad \hat{Q}=1-\hat{P}.
 \label{eq:s-gram-identity}
\end{equation}
At $\tau=i$, the bond source has $u_1=\delta\tau_2$,
$u_2=-\delta\tau_1$. Holomorphicity implies
$\hat{Q}\partial_{u_2}\Phi=i\hat{Q}\partial_{u_1}\Phi$, and its adiabatic curvature is
\[
 i\tr \hat{P}[\partial_{u_1}\hat{P},\partial_{u_2}\hat{P}]
 =-2\partial_{\bar\tau}\partial_\tau\ln\det G.
\]
With $C=-1$, $\rho=p/q$, and
$\langle f\rangle=q\int_{\rm MBZ}d^2k\,f/(2\pi)^2$, this proves
\begin{equation}
 \bar\eta_H\equiv\frac{\eta_H}{C\rho}
 =\left.\frac2p\left\langle
 \partial_{\bar\tau}\partial_\tau\ln\det G
 \right\rangle\right|_{\tau=i}.
 \label{eq:s-gram-viscosity}
\end{equation}
The momentum-space ideal condition and the $\tau$-holomorphic frame
express related properties of the sampled LLL. The step needed in
Eq.~\eqref{eq:s-gram-identity} is specifically $\partial_{\bar\tau}\Phi=0$;
momentum-space ideality alone does not determine a strain response.

\section{Poisson images and the continuum limit}

Poisson summation of the orbital overlap in
Eq.~\eqref{eq:s-theta-frame} gives
\begin{align}
 G_{rs}(\bm k;\tau)
 &=A\sum_{\ell,m\in\mathbb Z}(-1)^{qm\ell}
 e^{-x|m+\tau(s-r+p\ell)|^2/(2\tau_2)}
 e^{iq\ell k_x-i(q/p)mk_y-i\pi(q/p)m(r+s)},\nonumber\\
 A&=(2\phi\tau_2)^{-1/2},\qquad x=\frac\pi\phi.
 \label{eq:s-poisson}
\end{align}
The zero image is $G_0=A\mathbb I_p$. Writing $G=A(\mathbb I_p+\mathsf E)$
separates its continuum contribution,
\begin{equation}
 \frac2p\partial_{\bar\tau}\partial_\tau
 \langle\ln\det G\rangle
 =\frac1{4\tau_2^2}
 +\frac2p\partial_{\bar\tau}\partial_\tau
 \langle\ln\det(\mathbb I_p+\mathsf E)\rangle.
 \label{eq:s-sampling-correction}
\end{equation}
All derivatives hold material momentum and flux fixed. The second term
is the sampling correction, and the first yields $\bar\eta_H=1/4$.

To obtain its leading asymptotics at arbitrary rank $p$, group images
by $(m,N)$ with $N=s-r+p\ell$. Each image multiplies a unitary shift
matrix; opposite images are Hermitian conjugates. The traced MBZ
average enforces cancellation of the total image labels, so
\begin{equation}
 \langle\tr\mathsf E\rangle=0,\qquad
 \frac1p\langle\tr\mathsf E^2\rangle
 =\sum_{(m,N)\ne(0,0)}e^{-x|m+\tau N|^2/\tau_2}.
 \label{eq:s-image-pairing}
\end{equation}
Keeping the four shortest images in
$\tr\ln(1+\mathsf E)=\tr\mathsf E-\tr\mathsf E^2/2+\cdots$
gives, near $\tau=i$,
\begin{equation}
 \frac1p\langle\ln\det G\rangle
 =\ln A-e^{-x/\tau_2}-e^{-x|\tau|^2/\tau_2}
 +O(e^{-3x/2}).
 \label{eq:s-log-expansion}
\end{equation}
At the square point the Laplacian of each displayed exponential is
$(x^2-2x)e^{-x}$. Since
$2\partial_{\bar\tau}\partial_\tau
=(\partial_{\tau_1}^2+\partial_{\tau_2}^2)/2$, one obtains
\begin{equation}
 \bar\eta_H=\frac14-(x^2-2x)e^{-x}
 +O(x^2e^{-3x/2}).
 \label{eq:s-viscosity-asymptotic}
\end{equation}

\section{Straight-bond current and charge continuity}

We use the single-particle magnetic hopping operators in Eq.~\eqref{eq:km-translation}, with the background field denoted by $\bm A_0$ and the probe by $\bm A$. Coupling the probe along the straight bond gives
\begin{equation}
\hat H[A]=\sum_{\bm R,\bm d\ne0}t(\bm d)
 e^{-iBd_y(R_x+d_x/2)}
 \exp\!\left[-i\int_{-1/2}^{1/2}ds\,
 d_iA_i(\bm R+\bm d/2+s\bm d)\right]
 |\bm R\rangle\langle\bm R+\bm d|.
\label{eq:s-straight-peierls}
\end{equation}
The directed sum includes both bond orientations. We use
$A_i(\bm r)=\int d^2q\,A_i(\bm q)e^{i\bm q\cdot\bm r}/(2\pi)^2$.
The bond integral of a Fourier component is
$d_iA_i(\bm q)e^{i\bm q\cdot(\bm R+\bm d/2)}
\operatorname{sinc}(\bm q\cdot\bm d/2)$.
Functional differentiation therefore gives
\begin{align}
\hat J_i(\bm q)
&=i\sum_{\bm R,\bm d\ne0}d_i t(\bm d)e^{-iBd_y(R_x+d_x/2)}
 e^{-i\bm q\cdot(\bm R+\bm d/2)}
 \operatorname{sinc}(\bm q\cdot\bm d/2)
 |\bm R\rangle\langle\bm R+\bm d|\nonumber\\
&=i\sum_{\bm d\ne0}d_i t(\bm d)\operatorname{sinc}(\bm q\cdot\bm d/2)
 \hat N(\bm q/2)\hat{\mathcal W}_{\bm d}\hat N(\bm q/2),
\qquad \hat N(\bm q)=e^{-i\bm q\cdot\hat{\bm r}}.
\label{eq:s-current-definition}
\end{align}
The two density factors place the Fourier phase at the bond center;
the sinc factor averages the probe along the entire bond.
The definition obeys $\hat J_i(\bm q)^\dagger=\hat J_i(-\bm q)$.

To prove charge continuity, note that
\begin{equation}
[\hat N(\bm q),\hat{\mathcal W}_{\bm d}]
=2i\sin(\bm q\cdot\bm d/2)
\hat N(\bm q/2)\hat{\mathcal W}_{\bm d}\hat N(\bm q/2).
\label{eq:s-density-translation}
\end{equation}
Since $(\bm q\cdot\bm d)\operatorname{sinc}(\bm q\cdot\bm d/2)
=2\sin(\bm q\cdot\bm d/2)$, this yields
\begin{equation}
q_i\hat J_i(\bm q)=[\hat N(\bm q),\hat H],
\qquad
\dot{\hat N}(\bm q)=i[\hat H,\hat N(\bm q)]
=-iq_i\hat J_i(\bm q).
\label{eq:s-charge-ward}
\end{equation}
Second quantization replaces each $|\bm R\rangle\langle\bm R'|$
by $\hat c_{\bm R}^\dagger\hat c_{\bm R'}$ and preserves these
commutators. In particular,
\begin{equation}
\dot{\hat\rho}_{\bm q}
+iq_i\sum_{\bm R,\bm R'}
\hat c_{\bm R}^\dagger
\langle\bm R|\hat J_i(\bm q)|\bm R'\rangle
\hat c_{\bm R'}=0.
\end{equation}
This derivation applies on the infinite lattice at arbitrary $B$;
it also applies to an open sample with the bond sum restricted to its
sites. For rational flux, the same operators can be represented in
magnetic Bloch fibers for numerical evaluation. Rationality is a
condition for that finite-dimensional representation, not for the
current or its conservation law.

For the midpoint prescription, define the uniform velocity
$\hat v_i=\mathcal D_i\hat H=i[\hat H,\hat r_i]$ and
\begin{equation}
\hat J_i^{m}(\bm q)=\hat N(\bm q/2)\hat v_i\hat N(\bm q/2).
\label{eq:s-midpoint-current}
\end{equation}
The centered straight-bond operator is
\begin{equation}
\hat J_i^{\rm c}(\bm q)
=i\sum_{\bm d\ne0}d_i t(\bm d)
\operatorname{sinc}(\bm q\cdot\bm d/2)\hat{\mathcal W}_{\bm d},
\qquad
\hat J_i(\bm q)=\hat N(\bm q/2)\hat J_i^{\rm c}(\bm q)\hat N(\bm q/2).
\label{eq:s-centered-current}
\end{equation}
Using $\mathcal D_i\hat{\mathcal W}_{\bm d}=id_i\hat{\mathcal W}_{\bm d}$,
its small-$q$ expansion and the midpoint Ward residual are
\begin{align}
\hat J_i^{\rm c}(\bm q)
&=\hat v_i+\frac{q_jq_l}{24}\mathcal D_i\mathcal D_j\mathcal D_l\hat H+O(q^4),\nonumber\\
q_i\hat J_i^m(\bm q)-[\hat N(\bm q),\hat H]
&=-\hat N(\bm q/2)
\left[\frac{q_iq_jq_l}{24}\mathcal D_i\mathcal D_j\mathcal D_l\hat H+O(q^5)\right]
\hat N(\bm q/2).
\label{eq:s-midpoint-residual}
\end{align}
The endpoint factors are kept exact in this expression. The midpoint
operator is a specified smeared observable, but is not the conserved
source current for the site density. Different Peierls paths with the
same endpoints preserve Eq.~\eqref{eq:s-charge-ward}; the difference
between their currents is transverse. Length- and velocity-gauge
responses agree when the source, density, and electromagnetic contact
terms are carried through consistently.

For the quadratic continuum Hamiltonian $\hat H=\hat{\bm\pi}^2/(2m)$,
the commutator derivative gives
\begin{equation}
\mathcal D_i\hat H=\frac{\hat\pi_i}{m},\qquad
\mathcal D_j\mathcal D_i\hat H=\frac{\delta_{ij}}m,\qquad
\mathcal D_l\mathcal D_j\mathcal D_i\hat H=0.
\end{equation}
The higher-order corrections distinguishing the straight-source and
midpoint currents therefore vanish. The local current density
$\hat J_i(\bm x)=\{\hat\pi_i,\delta(\bm x-\hat{\bm r})\}/(2m)$ gives the same
Fourier current, since
\begin{equation}
\frac1{2m}\{\hat\pi_i,e^{-i\bm q\cdot\hat{\bm r}}\}
=\frac1m e^{-i\bm q\cdot\hat{\bm r}/2}\hat\pi_i e^{-i\bm q\cdot\hat{\bm r}/2}.
\end{equation}
This equality concerns the same quadratic continuum source; lattice
sampling of its occupied states alone does not imply equality of the
corresponding finite-wave-vector lattice currents.

\section{Response prescriptions}

The comparison uses the filled-manifold energy density
$\mathcal E(B,\kappa,\gamma)$ at $\nu=1$:
\begin{equation}
 \chi_B=-\left.\partial_B^2\mathcal E\right|_{\kappa,\gamma,\nu=1},
 \qquad R_\chi^{\rm LF}=-2\pi m_{\rm LF}\chi_B.
 \label{eq:s-chi-protocol}
\end{equation}
The density $\rho=B/(2\pi)$ varies along this fixed-envelope path.
Differentiating instead on the parent curve $\kappa=\pi-B/2$ gives
$\chi_{\rm parent}=\chi_B+\mathcal E_{B\kappa}
-\mathcal E_{\kappa\kappa}/4=0$, because
$\mathcal E_{\rm parent}=-B/(2\pi)$.
It therefore probes a different susceptibility.

The low-frequency mass follows from the uniform Peierls response. In expectation values below, each single-particle operator denotes its second-quantized fermion bilinear.
Write $\hat{H}[A_x]=\hat{H}_0-A_x\hat J_x+A_x^2\hat D_{xx}/2+\cdots$.
With $J_x^{mn}=\langle m|\hat J_x(0)|n\rangle$ denoting the single-particle current matrix elements, the retarded paramagnetic kernel and its insulating contact
cancellation are
\begin{equation}
 \Pi^R_{xx}(\omega)=\frac1{\mathcal A}
 \sum_{n\in\mathrm{occ},m\in\mathrm{emp}}
 \frac{2\Delta_{mn}|J_x^{mn}|^2}
 {(\omega+i0^+)^2-\Delta_{mn}^2},\qquad
 D\equiv\frac{\langle\hat D_{xx}\rangle}{\mathcal A}
 =-\Pi^R_{xx}(0).
 \label{eq:s-uniform-kernel}
\end{equation}
For $e^{-i\omega t}$, the physical response is
$\delta\langle\hat J_x\rangle/\mathcal A=-(D+\Pi^R_{xx})A_x$, whereas $E_x=i\omega A_x$.
Expanding the kernel therefore gives
\begin{equation}
 \sigma_{xx}(\omega)
 =-i\omega\frac{\rho m_{\rm LF}}{B^2}+O(\omega^3),\qquad
 m_{\rm LF}=\frac{2B^2}{\rho\mathcal A}
 \sum_{n\in\mathrm{occ},m\in\mathrm{emp}}\frac{|J_x^{mn}|^2}{\Delta_{mn}^3}.
 \label{eq:s-lf-mass}
\end{equation}
Here $n,m$ label the full single-particle spectrum; a magnetic Bloch decomposition, when available, resolves these sums into momenta and bands. This mass measures the uniform dynamical polarizability. On a lattice,
it is a response prescription, rather than a mass selected uniquely
by a Galilean Ward identity.

\section{Leading weak-flux correction}

\subsection{Magnetic translations and the effective Hamiltonian}

The relation between hopping and covariant momentum can be established
directly. For $\hat{\bm\pi}=-i\bm\nabla-\bm A_0$, introduce the continuously
translated wave function
\[
\psi_s(\bm r)=\bigl(e^{is\bm d\cdot\hat{\bm\pi}}\psi\bigr)(\bm r),
\qquad \psi_0(\bm r)=\psi(\bm r),
\]
where $s\in[0,1]$ parameterizes the displacement along the bond.
It obeys
$\partial_s\psi_s=(\bm d\cdot\bm\nabla-i\bm d\cdot\bm A_0)\psi_s$,
whose solution is
\begin{equation}
\psi_s(\bm r)=
\exp\!\left[-i\int_0^s dt\,
\bm d\cdot\bm A_0(\bm r+t\bm d)\right]
\psi(\bm r+s\bm d).
\label{eq:s-covariant-translation}
\end{equation}
At $s=1$, this reads the wave-function amplitude at $\bm r+\bm d$
and transports it to $\bm r$ with the straight-bond Peierls phase. In Landau gauge $\bm A_0=(0,Bx)$,
the phase is $e^{-iBd_y(x+d_x/2)}$, reproducing the
row-to-column hopping operator $\hat{\mathcal W}_{\bm d}$ of Eq.~\eqref{eq:km-translation}. Hence
\begin{align}
\hat{\mathcal W}_{\bm d}&=e^{i\bm d\cdot\hat{\bm\pi}},\nonumber\\
\hat{\mathcal W}_{\bm d}\hat{\mathcal W}_{\bm e}
&=e^{-iB(\bm d\times\bm e)/2}\hat{\mathcal W}_{\bm d+\bm e}.
\label{eq:s-weyl-map}
\end{align}
The second line follows from
$[\hat{\pi}_x,\hat{\pi}_y]=iB$ and the Baker--Campbell--Hausdorff formula.
Integer displacements reproduce the lattice hopping; this
representation does not assume continuous translation symmetry
of the lattice.

The hopping Hamiltonian and its commuting-momentum symbol are
therefore
\begin{equation}
\hat{H}=\sum_{\bm d\ne0}t(\bm d;\kappa)e^{i\bm d\cdot\hat{\bm\pi}},
\qquad
\varepsilon_\kappa(\bm k)
=\sum_{\bm d\ne0}t(\bm d;\kappa)e^{i\bm k\cdot\bm d},
\qquad
t(\bm d;\kappa)=\varsigma_{\bm d}e^{-\kappa d^2/2}.
\label{eq:s-hopping-symbol}
\end{equation}
Expanding the first expression fixes the operator ordering:
each commuting monomial in the second expression is replaced
by the fully symmetrized product of covariant momenta.
This is Weyl ordering. Here $\bm k$ is the commuting symbol variable, not a magnetic Bloch momentum; no rational-flux assumption enters this construction.

At the zero-field parent point $\kappa=\pi$, define
\[
M_{ab}=\sum_{m,n}m^an^b\varsigma_{mn}
e^{-\pi(m^2+n^2)/2}.
\]
The Perelomov identity with a holomorphic linear term reads,
for arbitrary $\lambda$,
\[
\sum_{m,n}\varsigma_{mn}
e^{-\pi(m^2+n^2)/2+\lambda(m+in)}=0.
\]
Its fourth derivative at $\lambda=0$, together with square
symmetry, gives $2M_{40}-6M_{22}=0$.
Likewise, differentiating the unimodular Gaussian identity
\[
\sum_{m,n}\varsigma_{mn}
e^{-\pi(e^t m^2+e^{-t}n^2)/2}=0
\]
twice at $t=0$ gives
$\pi(M_{40}-M_{22})=2M_{20}$.
Consequently,
\begin{equation}
\varepsilon_0(\bm k)
=-1+\alpha k^2+\beta(k^2)^2+O(k^6),
\qquad
\alpha=-\frac{M_{20}}2>0,
\qquad
\beta=\frac{M_{40}}{24}
=-\frac{\alpha}{4\pi}.
\label{eq:s-gaussian-moments}
\end{equation}
The constant $-1$ results from excluding the
on-site term from the vanishing Perelomov sum.

Along $\kappa=\pi-B/2$, the hopping acquires the factor
$e^{Bd^2/4}$. Since
$\nabla_k^2e^{i\bm k\cdot\bm d}=-d^2e^{i\bm k\cdot\bm d}$,
the corresponding symbols satisfy
$\varepsilon_\phi=e^{-B\nabla_k^2/4}\varepsilon_0$.
Using $\nabla_k^2k^2=4$ and
$\nabla_k^2(k^2)^2=16k^2$ gives
\begin{equation}
\varepsilon_\phi(\bm k)
=-1+\alpha(k^2-B)
+\beta[(k^2)^2-4Bk^2+2B^2]
+O(B^3),
\label{eq:s-parent-symbol}
\end{equation}
where combined order counts $k^2$ as $O(B)$.

To recover the Hamiltonian, define
$\hat{\bm\pi}^2=\hat{\pi}_x^2+\hat{\pi}_y^2$.
The quadratic term becomes $\hat{\bm\pi}^2$, whereas the fully symmetrized
quartic term becomes $(\hat{\bm\pi}^2)^2+B^2$.
Explicitly, the moment identities give
\begin{equation}
\frac1{24}\sum_{\bm d\ne0}t(\bm d;\pi)
(\bm d\cdot\hat{\bm\pi})^4
=\beta\bigl[(\hat{\bm\pi}^2)^2+B^2\bigr].
\label{eq:s-quartic-ordering}
\end{equation}
The additional $B^2$ follows from commuting $\hat{\pi}_x$ and $\hat{\pi}_y$
when collecting the symmetrized products. Thus
\begin{equation}
\hat{H}_{\rm eff}=-1+\alpha(\hat{\bm\pi}^2-B)
+\beta(\hat{\bm\pi}^2-B)(\hat{\bm\pi}^2-3B)+O(B^3).
\label{eq:s-effective-hamiltonian}
\end{equation}
On fixed low Landau levels, $\hat{\bm\pi}^2$ has eigenvalues $B(2n+1)$
and is therefore $O(B)$. The quartic term provides the leading
relative correction to the quadratic theory. Equation~\eqref{eq:s-effective-hamiltonian}
follows the parent curve; the field derivatives defining $\chi_B$
instead hold the envelope fixed, as specified below.

\subsection{Conductivity, susceptibility, and their comparison}

Using $\hat{\bm\pi}^2|n\rangle=B(2n+1)|n\rangle$, the excitation gaps
from the occupied $n=0$ state are
\begin{equation}
\Delta_n=2\alpha Bn+4\beta B^2n(n-1)+O(B^3).
\label{eq:s-oscillator-data}
\end{equation}
In particular, $\Delta_1=2\alpha B+O(B^3)$ and
$\Delta_2=4\alpha B+8\beta B^2+O(B^3)$.
Within the quartic truncation, the uniform velocity and centered
straight-source current operators of Eq.~\eqref{eq:s-centered-current}, represented in the oscillator basis, are
\begin{align}
\hat{v}_i&=2(\alpha-4\beta B)\hat{\pi}_i
+2\beta \{\hat{\pi}_i,\hat{\bm\pi}^2\},\nonumber\\
\hat J_x^{\rm c}(q_x)&=\hat{v}_x+\beta q_x^2\hat{\pi}_x,
\qquad
\hat J_y^{\rm c}(q_x)=\hat{v}_y+\frac{\beta}{3}q_x^2\hat{\pi}_y.
\label{eq:s-effective-vertices}
\end{align}
The omitted Hamiltonian terms contribute $O(B^{5/2})$ to
low-level matrix elements of the uniform velocity.

To evaluate the finite-wave-vector response, distinguish the centered
current $\hat J_y^{\rm c}(q_x)$ from the current operator carrying momentum
$q_x$. In the effective Landau-level description, write
$\hat x=\hat X+\hat{r}_x^{\rm cyc}$, where $\hat X$ is the guiding-center coordinate and
$\hat{r}_x^{\rm cyc}=-\hat{\pi}_y/B$. The opposite guiding-center Fourier factors
cancel in the correlator of a uniformly filled level. The remaining
cyclotron operators are
\begin{equation}
\hat{J}_y(q_x)=e^{-iq_x\hat{r}_x^{\rm cyc}/2}\hat J_y^{\rm c}(q_x)
e^{-iq_x\hat{r}_x^{\rm cyc}/2},
\qquad
\hat{N}(-q_x)=e^{iq_x\hat{r}_x^{\rm cyc}}.
\label{eq:s-cyclotron-vertices}
\end{equation}
Here $\hat N$ denotes the cyclotron part of the site-density operator, and the two exponential factors
in $\hat{J}_y$ restore the Fourier phase of the bond center.

Define the static retarded density--current correlator per particle by
its spectral representation,
\begin{align}
\mathcal R_{JN}(0;q_x)
=-\sum_{n\ge1}\frac{1}{\Delta_n}\Big[
&\langle0|\hat{J}_y(q_x)|n\rangle
 \langle n|\hat{N}(-q_x)|0\rangle \nonumber\\
+{}&\langle0|\hat{N}(-q_x)|n\rangle
 \langle n|\hat{J}_y(q_x)|0\rangle\Big].
\label{eq:s-density-current-correlator}
\end{align}
The sum runs over transitions from the occupied $n=0$ level to
empty levels. Expanding the density and current operators shows that
the $n=1$ channel contributes at orders $q_x$ and $q_x^3$, while the
$n=2$ channel first contributes at $q_x^3$. Higher levels enter at
$q_x^5$ or beyond. Within the quartic theory, the two contributions are
\begin{align}
\mathcal R_{JN}^{(1)}(0;q_x)
&=\frac{iq_x}{B}
-\frac{iq_x^3}{B^2}
\left(1+\frac{4\beta B}{3\alpha}\right)+O(q_x^5),
\nonumber\\
\mathcal R_{JN}^{(2)}(0;q_x)
&=\frac{iq_x^3}{4B^2}+O(q_x^5).
\label{eq:s-two-transitions}
\end{align}
The physical correlator per area is $\rho\mathcal R_{JN}$.
With the Hall orientation used here,
\begin{equation}
\sigma_H(q_x)
=-\frac{i\rho}{q_x}\mathcal R_{JN}(0;q_x)
=\frac{\rho}{B}
\left[1+C_2\frac{q_x^2}{B}+O(q_x^4)\right].
\label{eq:s-density-current-conductivity}
\end{equation}
Adding the two channels therefore gives
\begin{equation}
C_2=-\frac34-\frac{4\beta}{3\alpha}B+O(B^2)
=-\frac34+\frac23\phi+O(\phi^2).
\label{eq:s-conductivity-asymptotic}
\end{equation}

For the susceptibility, the distinction between the reference field
and the varied field must be retained. Fix the envelope at
$\kappa_B=\pi-B/2$, and denote the varied physical field by
$\mathcal B$. The hopping coefficients then retain their reference
values; only the covariant-momentum algebra changes to
$[\hat{\pi}_x,\hat{\pi}_y]=i\mathcal B$.
Weyl ordering of the fixed symbol in Eq.~\eqref{eq:s-parent-symbol}
gives the occupied energy per particle and its density-weighted value,
\begin{align}
E_{\rm occ}(\mathcal B;\kappa_B)
&=-1+\alpha(\mathcal B-B)
+2\beta(\mathcal B-B)^2+O_3,\nonumber\\
\mathcal E(\mathcal B;\kappa_B)
&=\frac{\mathcal B}{2\pi}E_{\rm occ}(\mathcal B;\kappa_B).
\label{eq:s-fixed-envelope-energy}
\end{align}
Here $O_3$ denotes terms of combined degree at least three in
$B$ and $\mathcal B$ in the smooth weak-field expansion.
Differentiating with respect to $\mathcal B$ at fixed $B$ and then
setting $\mathcal B=B$ yields
\begin{align}
\left.\partial_{\mathcal B}^2\mathcal E
\right|_{\mathcal B=B,\,\kappa_B}
&=\frac{2\alpha+4\beta B}{2\pi}+O(B^2),\nonumber\\
R_\chi^{\rm LF}=-2\pi m_{\rm LF}\chi_B
&=1+\frac{2\beta}{\alpha}B+O(B^2)
=1-\phi+O(\phi^2).
\label{eq:s-susceptibility-asymptotic}
\end{align}
Substituting $\mathcal B=B$ before differentiating would instead
follow the parent energy $-B/(2\pi)$ and compute a different
susceptibility.

The exact Gram-matrix result supplies $\bar\eta_H=1/4$ to algebraic
order. Combining the three responses gives
\begin{equation}
\Delta
=C_2-\bar\eta_H+R_\chi^{\rm LF}
=\frac{2\beta}{3\alpha}B+O(B^2)
=-\frac{B}{6\pi}+O(B^2)
=-\frac\phi3+O(\phi^2).
\label{eq:s-hoyos-asymptotic}
\end{equation}
Only the constant and linear terms are controlled by this truncation.
The response remainder assumes an isolated quadratic minimum,
relevant gaps bounded below by a positive multiple of $B$, and control
of the high-energy occupied--empty resolvent. The $q_x,\omega\to0$
limits are taken at fixed $B>0$ before $\phi\to0$.

These responses retain different spectral and source information.
The bond Hall viscosity is fixed by the strain geometry of the
occupied manifold; in the isotropic Landau-level description,
uniform shear probes the quadrupolar channel, with the excitation
energies canceling against the stress matrix elements in the
adiabatic response. The nonlocal Hall conductivity samples both
dipolar and quadrupolar transitions, retaining their relative
energies and the finite-wave-vector current matrix elements. The
fixed-envelope susceptibility instead measures the field dependence
of the occupied energy and degeneracy, and therefore probes the
Hamiltonian away from the parent curve. Different transition
channels do not themselves obstruct a Hoyos relation: Galilean
Ward identities tie them together in the continuum. On the lattice,
such a reduction requires additional operator constraints and
does not follow from the sampled-LLL structure alone.

\end{document}